\documentstyle[emulateapj,apjfonts,epsfig]{article}

\newcommand{\avz}{\langle z\rangle}
\newcommand{\be}{\begin{equation}}
\newcommand{\ee}{\end{equation}}

\begin{document}

\submitted{Submitted to The Astrophysical Journal}

\title{Hydrostatic Expansion and Spin Changes During Type I X-Ray
Bursts}

\author{Andrew Cumming, Sharon M. Morsink\altaffilmark{1}, 
Lars Bildsten\altaffilmark{2}, 
John L. Friedman\altaffilmark{3}, \& Daniel
E. Holz}

\affil{Institute for Theoretical Physics, Kohn Hall,
University of California, Santa Barbara, CA 93106}

\altaffiltext{1}{Permanent address: Department of Physics, University of Alberta,
Edmonton, AB, T6G 2J1, Canada}
\altaffiltext{2}{Department of Physics,
University of California, Santa Barbara, CA 93106}
\altaffiltext{3}{Permanent address: Department of Physics, University of
Wisconsin-Milwaukee, P.O. Box 413, Milwaukee, WI 53201}

\begin{abstract}
We present calculations of the spin-down of a neutron star atmosphere
due to hydrostatic expansion during a Type I X-ray burst. We show that
Cumming \& Bildsten incorrectly calculated the change in the moment of
inertia of the atmosphere during a Type I burst, resulting in a factor
of two overestimation of the magnitude of the spin-down for rigidly
rotating atmospheres. We derive the angular momentum conservation law
in general relativity, both analytically for the case of slow
rotation, and numerically for rapidly-rotating stars. We show that
contrary to the claims of Heyl, but in agreement with Abramowicz and
coworkers, general relativity has a small effect on the angular
momentum conservation law, at the level of 5--10\%. We show how to
rescale our fiducial results to different neutron star masses,
rotation rates and equations of state, and present some detailed
rotational profiles. Comparing our results with recent observations of
large frequency shifts in MXB 1658-298 and 4U 1916-053, we find that
the spin-down expected if the atmosphere rotates rigidly is a factor
of two to three less than the observed values. If differential
rotation is allowed to persist, we find that the upper layers of the
atmosphere do spin down by an amount comparable to or greater than the
observed values. However, there is no compelling reason to expect the
observed spin frequency to be that of only the outermost layers of the
atmosphere. We conclude that hydrostatic expansion and angular
momentum conservation alone cannot account for the largest frequency
shifts observed during Type I bursts.
\end{abstract}


%-------------------------------------------------------------------------

\section{Introduction}\label{sec:Intro}

Type I X-ray bursts are thermonuclear flashes on the surface of
weakly-magnetic ($B\lesssim 10^{10}\ {\rm G}$) accreting neutron stars
in low mass X-ray binaries (LMXBs) (Lewin, van Paradijs, \& Taam 1995;
Bildsten 1998). Observations with the Rossi X-ray Timing Explorer
(RXTE) have revealed nearly-coherent oscillations during Type I X-ray
bursts from ten LMXBs, with frequencies in the range 270--620 Hz (for
reviews, see van der Klis 2000; Strohmayer 2001). These oscillations
are interpreted as rotational modulation of surface brightness
asymmetries, thus providing a direct measurement of the neutron star
spin frequency (Strohmayer et al.~1996). The inferred rotation periods
of a few milliseconds support the idea that the neutron stars in LMXBs
are the progenitors of the millisecond radio pulsars (see Bhattacharya
1995 for a review).

The large modulation amplitudes, high coherence, and frequency
stability provide convincing evidence that the burst oscillation
frequency is related to the neutron star spin frequency (Strohmayer
2001 and references therein). However, many puzzles remain. Open
questions include (i) the origin of the asymmetry on the surface of
the neutron star, particularly during the fading tail of the burst,
when the burning is thought to have spread over the whole surface;
(ii) why oscillations are seen in some bursts but not others, and in
some sources but not others (e.g., Muno et al.~2001); and (iii)
whether the oscillation frequency is the spin frequency or twice the
spin frequency, motivated by comparisons of the burst oscillation
frequency to the difference between the frequencies of the kHz
quasi-periodic oscillations (QPOs) seen in the persistent emission
(van der Klis 2000).

The oscillation frequency increases by a few Hz during the burst. To
explain this, Strohmayer et al.~(1997) proposed that the burning shell
decouples from the neutron star, and undergoes spin changes due to
angular momentum conservation as it expands and contracts. Cumming \&
Bildsten (2000, hereafter CB) studied the rotational evolution of the
neutron star atmosphere during a burst. Rather than addressing the
complex problems of how the burning front spreads over the neutron
star surface during the burst rise, or what causes the asymmetry at
late times, CB asked what could be learned from detailed models of the
vertical structure of the atmosphere. They computed the vertical
hydrostatic expansion during the burst, and the resulting spin
changes. They noted that to see a single, coherent frequency requires
either the atmosphere be rigidly rotating, or the burning region to be
vertically thin, so that the differential rotation across it is small
(see also Miller 2000). In order to compare the theoretical results
with observations, CB assumed that the burning shell rotates
rigidly. They found rough agreement between the calculated spin down
and observed values.

The composition of the burning shell affects the amount of
expansion. For accretion rates relevant to X-ray bursters, the
accreted hydrogen (H) burns via the hot CNO cycle as matter
accumulates on the neutron star surface. The thermally unstable helium
(He) burning which leads to the Type I X-ray burst may occur either
before or after the hydrogen has been consumed (see Bildsten 1998 for
a review), leading to either mixed hydrogen/helium or pure helium
ignitions. The presence of a substantial amount of hydrogen leads to a
factor of two greater expansion because of the lower mean molecular
weight ($\mu=4/3$ for pure He; $\mu\approx 0.6$ for solar
composition). CB noted that the largest frequency shifts observed
(fractional frequency shifts $\Delta\nu/\nu\approx 0.8$\%) were
difficult to achieve with pure He ignitions, requiring temperatures at
the base of the burning shell extremely close to the limiting value
from radiation pressure at the start of the burst. However, they also
noted that no burst oscillations have been seen during Type I X-ray
bursts with long ($\gtrsim 20\ {\rm s}$) cooling tails, characteristic
of hydrogen burning (which is limited by beta decays during the rp
process; Bildsten 1998), perhaps indicating a preponderance of He
bursts amongst those which show oscillations. Thus they suggested that
understanding the largest frequency shifts observed as being due to
hydrostatic expansion together with rigid rotation might prove
problematic.

Recently, two sources have shown extremely large frequency
shifts. Wijnands, Strohmayer, \& Franco (2001) found an increase of
$\approx 5\ {\rm Hz}$ in the 567 Hz burst oscillation from
MXB~1658-298. In addition to being a large frequency shift
($\Delta\nu/\nu\approx 0.9$\%), the spin evolution in this burst was
unusual, with a rapid frequency increase occuring several seconds into
the burst tail. This frequency shift is a little larger than the
rigidly-rotating mixed H/He models of CB. Galloway et al.~(2001)
discovered a 270 Hz burst oscillation from the dipping source
4U~1916-053. During 4 seconds in the burst decay, the frequency
increased by 3.6 Hz, a fractional shift of 1.3\%, by far the largest
fractional frequency shift observed so far. Comparing with the results
of CB, Galloway et al.~(2001) pointed out that the expansion implied
by this frequency shift is larger than expected if the atmosphere
rotates rigidly, especially given that the measured peak flux of the
X-ray burst is sub-Eddington ($\approx 0.5\ {\rm F_{\rm Edd}}$), and
that the orbital period of this source ($\approx 50\ {\rm mins}$;
Walter et al.~1982; White \& Swank 1982; Grindlay et al.~1988) implies
a hydrogen-poor companion (Nelson, Rappaport, \& Joss 1986) and
therefore a small fraction of hydrogen present at ignition.

In this paper, we present new calculations of the hydrostatic
expansion and resulting spin-down of the neutron star atmosphere. In
\S 2, we describe an error in CB's calculation of the change in the
moment of inertia of the burning shell, and show that the values of
spin-down calculated by CB are a factor of two too large. Comparing
the new results with observations, we show that, even when hydrogen is
present at ignition, {\it the theoretical values of spin-down assuming
rigid rotation are a factor of three less than the largest observed
frequency shifts}. We then go on to consider different neutron star
masses and equations of state. In \S 3, we derive the angular momentum
conservation law in general relativity, presenting analytic results
for slow rotation, and numerical calculations of rapidly-rotating
stars. We compare our results with recent work by Heyl (2000) and
Abramowicz, Kluzniak, \& Lasota (2001). In \S 4, we show how to
include general relativity and rapid rotation in the equations
describing the structure of the atmosphere. We then rescale our
fiducial results of \S 2 to different neutron star masses and
equations of state, before presenting some detailed models and
rotational profiles. We again show that the expected spin-down due to
angular momentum conservation is smaller than the observed frequency
drifts if the atmosphere rotates rigidly. We discuss the implications
of our results in \S 5.

\section{Calculation of Moment of Inertia and Comparison to Observations}

In this section, we first show that CB made an error of a factor of
two in their calculation of the change in the moment of inertia during
a Type I X-ray burst, and then compare the new spin down calculations
with observations.

Consider a spherical shell at the surface of the star. (The arguments
of this section are unaffected by including the centrifugal
distortion of the isobaric surfaces.) The mass of the shell is
\begin{equation}
M=\int dm=\int 4\pi r^2 \rho\ dr,
\end{equation}
and its moment of inertia is
\begin{equation}\label{eq:I}
I={2\over 3}\int r^2\ dm={8\pi\over 3}\int r^4 \rho\ dr,
\end{equation}
where the factor of $2/3$ comes from integrating over
angles. Following CB, we adopt a plane-parallel approximation for the
thin atmosphere, and write $r=R+z$, keeping only first order terms in
$z/R$. Changing integration variables from radius $r$ to column depth
$y$, where $dy=-\rho\,dr$, equation (\ref{eq:I}) reduces to CB's
equation (19) for the moment of inertia. If the angular momentum of
the atmosphere, $I\Omega$, is conserved, the spin-down is given by
comparing its moment of inertia before the burst with that during the
burst, $\Delta\Omega/\Omega=-\Delta I/I$.

The error in CB's calculation of the change in moment of inertia arises
because they assume that the total column depth of the atmosphere is
constant during the burst. We now show that this is not the case when
the mass of the layer is conserved. We write the mass of the shell as
\begin{equation}
M=4\pi R^2\int\left[1+{2z(y)\over R}\right]\ dy=4\pi
R^2\Delta y\left(1+{2\avz\over R}\right),
\end{equation}
where $\Delta y$ is the total column depth, and we define a
mass-weighted thickness $\avz=\int z(y) dy/\Delta y$. If the vertical
extent of the layer $\avz$ changes, so must its total column depth
$\Delta y$ if the total mass $M$ is conserved. Physically, a thin
spherical shell has a greater surface area if it moves radially
outwards, and so must have a lower column depth in order to conserve
its total mass. By taking the column depth at the base during the
burst to be the same as that before the burst, CB unwittingly
increased the mass of the layer during the burst, and so overestimated
the change in its moment of inertia.

Rather than including the change in the column depth of the layer
explicitly, we write the mass element in equation (\ref{eq:I}) as
$dm=4\pi R^2\ dy$, so that there is a unique correspondence between
mass and column depth. The moment of inertia is then
\begin{equation}\label{eq:right}
I={8\pi R^4\over 3}\int_{y_t}^{y_b} \left[1+{2z(y)\over R}\right]\ dy,
\end{equation}
where the integration limits are now held fixed during the burst,
thereby keeping the mass of the layer constant, $M=4\pi R^2\Delta
y=4\pi R^2(y_b-y_t)$. We have checked the validity of this expression
by integrating the full spherical equations for this problem. In this
paper, we adopt a plane-parallel approximation because it allows us to
straightforwardly incorporate the general relativistic angular
momentum conservation law.

The spin changes during the burst result from changes in the vertical
extent of the atmosphere, which are described by the second term of
equation (\ref{eq:right}). Equation (19) of CB has a factor $4z/R$
rather than $2z/R$, so that the spin-down calculated by CB is too
large by a factor of 2. Specifically, both the spin-down of the
convection zone shown in Figures 2 and 6 of CB, and the spin-down of a
rigidly-rotating atmosphere shown in Figure 13 of CB, should be
reduced by a factor of 2. The spin-down of the radiative layers shown
in Figures 2, 4 and 6 of CB is, however, calculated correctly, since
in that case CB assumed $r^2\Omega$ was constant for each spherical
shell.

\begin{figure}
\epsscale{1.0}\plotone{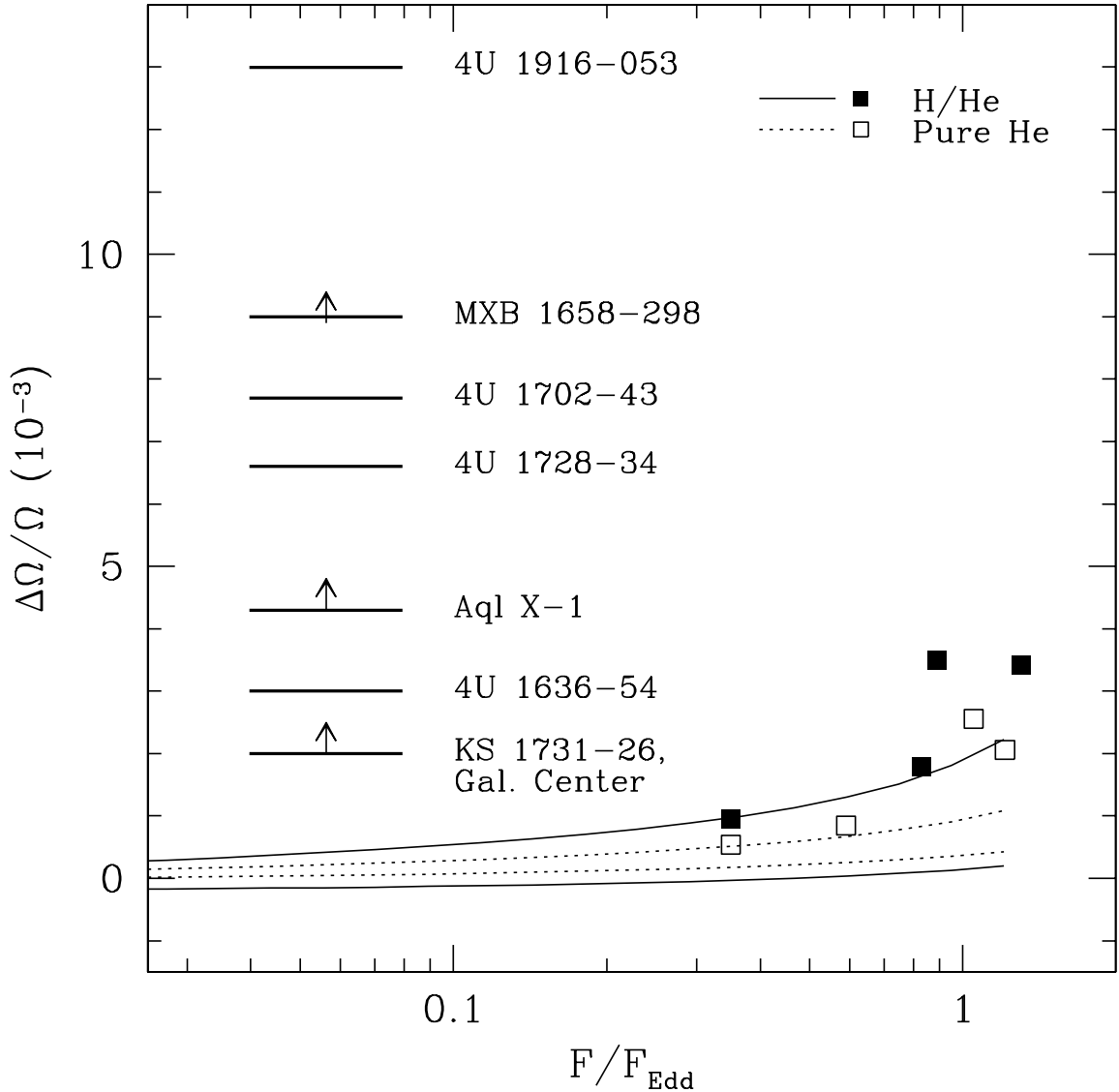}\epsscale{1.0}
\caption{The spin evolution of the atmosphere for convective ({\it
squares}) and radiative models ({\it lines}), assuming rigid rotation
is maintained throughout the atmosphere. We plot the fractional spin
frequency change of the burning shell, $\Delta\Omega/\Omega$, such
that $\Delta\Omega/\Omega>0$ indicates spin down. We show results for
both mixed H/He ignitions ({\it solid lines, solid squares}) and pure
He ignitions ({\it dotted lines, open squares}). For the convective
models, the composition is the same as before ignition. For the
radiative models, the upper curve is for a pre-burst composition; the
lower curve is for a composition of $^{56}$Ni (pure He) or $^{76}$Kr
(mixed H/He). We indicate the observed frequency shifts by horizontal
bars. These observations are summarized in Table 1 of CB; in addition,
we include the frequency shifts recently reported for 4U~1916-053
(Galloway et al.~2001) and MXB~1658-298 (Wijnands et al.~2001). For
those bursts in which the oscillation frequency was seen only in the
tail, we plot the $\Delta\Omega/\Omega$ value as a lower limit. This
Figure is an update of Figure 13 of CB.\label{fig:newsumm}}
\end{figure}

Figure \ref{fig:newsumm} shows the spin-down calculated using equation
(\ref{eq:right}) for the moment of inertia, presuming the whole
atmosphere rotates rigidly during the burst. This is an updated
version of Figure 13 of CB, and detailed discussion of the theoretical
models may be found in that paper. The calculated spin down is exactly
a factor of two less than found by CB. We take $g=1.9\times 10^{14}\
{\rm cm\ s^{-2}}$, the Newtonian gravity for a 1.4 $M_\odot$, 10 km
star. We consider two pre-burst models, mixed H/He ignition at $\dot
m=0.1\ \dot m_{\rm Edd}$ (solid lines and squares), and pure He
ignition at $\dot m=0.015\ \dot m_{\rm Edd}$ (dotted lines and open
squares).

Figure \ref{fig:newsumm} shows results for both convective and
radiative models. For the radiative models, we show the spin down as a
function of flux, since the radiative solution is uniquely determined
by the flux. The convective models depend both on the flux and the
extent of the convection zone; we plot the spin down resulting from
the convective models given in Table 3 of CB. The convective models
have a composition profile the same as that prior to the burst. The
radiative models have either a pre-burst composition (upper solid and
dotted curves) or a composition of heavy elements (lower solid and
dotted curves), either $^{56}$Ni for pure He burning, or $^{76}$Kr for
mixed H/He burning (we choose $^{76}$Kr as representative of the
products of rp process H burning; see Schatz et al.~1998; Koike et
al.~1999; Schatz et al.~2001). As described in CB, the radiative and
convective models with a pre-burst composition are appropriate for the
burst rise, whereas the radiative atmospheres composed of heavy
elements are appropriate for the cooling tail of bursts. Convective
models spin down more than radiative models because the steeper
temperature gradient in the convection zone leads to a larger base
temperature for a given flux.

On the left of Figure 1, we plot the observed frequency drifts as
horizontal bars. A summary of these observations may be found in CB,
Table 1; in addition, we include the frequency shifts recently
reported for 4U~1916-053 (Galloway et al.~2001) and MXB~1658-298
(Wijnands et al.~2001). For each source, we plot the largest observed
frequency shift. We plot as upper limits those cases for which a
frequency was only seen in the burst tail. The largest observed
frequency shift is larger than the theoretical values by at least a
factor of 3.

\section{Angular Momentum Conservation in General Relativity}

In this section, we derive the angular momentum conservation law in
general relativity. Our approach is to calculate the conserved angular
momentum of a particle moving on a circular path (in the Appendix, we
show that the same result is obtained starting with the conservation
equations for a fluid). We show analytic results for slowly rotating
stars, and present numerical calculations of rapidly rotating
stars. We compare our results with recent calculations by Heyl (2000)
and Abramowicz et al.~(2001).

In Newtonian physics, the specific angular momentum of a particle is
given by $r^2\Omega$, where $r$ is the distance from the rotation
axis. If the particle's motion conserves angular momentum, the change
in its spin frequency with radius is given by $d\ln\Omega/d\ln
r=-2$. In this section, we derive the equivalent expression in general
relativity. We start by discussing the angular momentum of a particle
moving in the Schwarzschild metric. We then derive the conserved
angular momentum in the slow rotation approximation, i.e.~neglecting
terms of order $(\Omega/\Omega_K)^2$, where $\Omega_K^2=GM/R^3$,
allowing us to write an analytic expression for the angular momentum
conservation law. However, since $(\Omega/\Omega_K)^2$ may be quite
large (for example, $(\Omega/\Omega_K)^2=0.26$ for a star with $R=15\
{\rm km}$ and $M=1.4 M_\odot$ spinning at $600\ {\rm Hz}$), we
conclude this section with detailed numerical calculations of rapidly
rotating models.

Consider first a spherical star, for which the metric is
\begin{equation}
ds^2=-e^\nu dt^2+e^\lambda dr^2+r^2\left(d\theta^2+\sin^2\theta d\phi^2\right),
\end{equation}
where outside the star the metric functions $\lambda$ and $\nu$ are
given by
\begin{equation}
e^\nu=e^{-\lambda}=1-{2M\over r}
\end{equation}
where $M$ is the total gravitational mass. Note that in this paper we
write the ratio $GM/rc^2$ as $M/r$ for clarity, but otherwise
explicitly write in the constants $G$ and $c$. A particle moving in a
circular path has four-velocity (e.g., Hartle 1970)
\begin{equation}\label{eq:fourvel}
u^\alpha=u^t\left(t^\alpha+\Omega(r)\phi^\alpha\right),
\end{equation}
where $\Omega(r)$ is the angular velocity measured by an observer at
infinity, $u^t$ is the redshift factor between the particle and the
observer at infinity, and $t^\alpha$ and $\phi^\alpha$ are Killing
vectors of the spacetime. The redshift factor is
\begin{equation}\label{eq:redshift}
u^t=e^{-\nu/2}=\left(1-{2M\over r}\right)^{-1/2},
\end{equation}
where we choose the standard normalization $g_{\alpha\beta}u^\alpha
u^\beta=-1$. Associated with the Killing vectors $t^\alpha$ and
$\phi^\alpha$ are the conserved quantities $E=-u^\alpha t_\alpha$ and
$L=u^\alpha \phi_\alpha$, the specific energy and angular
momentum. Using equations (\ref{eq:fourvel}) and (\ref{eq:redshift}),
we obtain (see, e.g. Schutz 1990,  \S 7.4)
\begin{equation}
E=\left(1-{2M\over r}\right)^{1/2},
\end{equation}
and
\begin{equation}\label{eq:L}
L=r^2\sin^2\theta\ \Omega\left(1-{2M\over r}\right)^{-1/2}.
\end{equation}
Differentiating equation (\ref{eq:L}) and evaluating it at the surface
of the star, we find
\begin{equation}
{d\ln\Omega\over d\ln r}=-2\left(1-{5M\over 2R}\right)
\left(1-{2M\over R}\right)^{-1}
\end{equation}
when a particle moves such that angular momentum is conserved. In the
limit $M\ll R$, this formula reduces to the Newtonian result.

It is simple to extend this analysis to the case of slow rotation, for
which the metric is (Hartle 1967)
\begin{equation}
ds^2=-e^\nu dt^2+e^\lambda dr^2+r^2(d\theta^2+\sin^2\theta d\phi^2)
-2\omega r^2\sin^2\theta\ d\phi dt,
\end{equation}
where $\lambda$ and $\nu$ have the same solutions as the spherical
case outside the star, and where $\omega=2J/r^3$ outside the star ($J$
is the total angular momentum of the star). Again solving for the
conserved angular momentum, we find
\begin{equation}\label{eq:slowL}
L=r^2\sin^2\theta\ \left(\Omega-\omega\right)\left(1-{2M\over
r}\right)^{-1/2},
\end{equation}
giving
\begin{equation}\label{eq:slow}
{d\ln\Omega\over d\ln r}=-2\left[1-{5M\over 2R}+{I\over
R^3}\left(1-{M\over R}\right)\right] \left(1-{2M\over R}\right)^{-1},
\end{equation}
at the surface of the star, where we presume the particle is initially
rotating with the spin of the star, and write the moment of inertia
$I=J/\Omega$.

For the rapidly-rotating case, we consider the general metric for an
axisymmetric stationary star,
\begin{equation}
ds^2 = - e^{\gamma + \rho} dt^2 
        + e^{\gamma - \rho} \bar{r}^2 \sin ^2 \theta \left(
                d\phi - \omega dt \right)^2 
        + e^{2 \alpha} \left( d\bar{r}^2 + \bar{r}^2 d\theta^2\right),
\label{rapid}
\end{equation}
where the metric potentials $\rho, \gamma, \alpha$ and $\omega$ depend
only on the coordinates $\bar{r}$ and $\theta$.  The coordinate
$\bar{r}$ is related to the Schwarzschild coordinate $r$ by $r =
\bar{r} \exp(\frac12(\gamma-\rho))$, so that $2 \pi r \sin \theta$ is
the circumference of a circle with constant $\bar{r}$ and $\theta$.
For further details about the interpretation of these coordinates and
potentials, see Friedman, Ipser \& Parker (1986) and Morsink \& Stella
(1999). Given an equation of state, a mass and a rotation rate, the
metric (\ref{rapid}) can be solved numerically. We use a code written
by N.~Stergioulas\footnote{Code publicly available at {\tt
http://www.gravity.phys.uwm.edu/rns/}}, which assumes rigid rotation
and is based on methods developed by Komatsu, Eriguchi \& Hachisu
(1989) and Cook, Shapiro \& Teukolsky (1994).

We present the properties of some rapidly-rotating models in Table
1. We consider two different equations of state (EOS), EOS L which is
stiff and EOS APR which is softer.  EOS L (Pandharipande \& Smith
1975) is one of the stiffest EOS in the Arnett \& Bowers (1977)
catalogue. EOS APR is the model A18+$\delta v$+UIX* computed by Akmal,
Pandharipande \& Ravenhall (1998) which uses modern nucleon scattering
data and first order special relativistic corrections.  For each EOS,
we consider two different masses, $M=1.4\ M_\odot$ and $2.0\ {\rm
M_\odot}$, and two spin frequencies, $\nu=300\ {\rm Hz}$ and $600\
{\rm Hz}$. The last column in Table 1 gives the break-up spin
frequency in each case.

We work in the equatorial plane, and do not consider the variations of
the metric potentials with latitude. The specific angular momentum is
\begin{equation}\label{eq:fastL}
L = \frac{ v r}{\sqrt{1-v^2}},
\end{equation}
(see also Morsink \& Stella 1999) where the three-velocity of a
corotating particle as measured by a zero angular momentum observer is
given by
\begin{equation}\label{eq:v}
v = r \; e^{-(\gamma + \rho)/2}\left[ \Omega(r) - \omega(r)\right].
\end{equation}
Again holding $L$ constant as we vary the Schwarzschild coordinate
radius $r$, we find
\begin{equation}\label{eq:fast}
{d\ln\Omega\over d\ln r}= -2\left[\left(1 - \frac{v^2}{2} -
\frac{R}{4}\left(\gamma_r + \rho_r\right)\right) \left(1-\frac{\omega}{\Omega}\right)-\frac{R}{2\Omega} \omega_r \right],
\end{equation}
where $R$ is the equatorial radius of the star, subscripts denote
partial differentiation and all quantities are evaluated at the
equator. In the slow rotation limit, equations (\ref{eq:fastL}) and
(\ref{eq:fast}) reduce to equations (\ref{eq:slowL}) and
(\ref{eq:slow}).

Heyl (2000) recently calculated $d\ln\Omega/d\ln r$, with a rather
different result (compare our eq.~[\ref{eq:slow}] with his
eq.~[10]). However, Abramowicz et al.~(2001) point out a sign error in
Heyl's calculation, and, more importantly, that Heyl assumes that the
quantity $L/E$ is conserved, as, for example, a particle in orbit
(Abramowicz \& Prasanna 1990). However, the energy $E$ of a fluid
element in the atmosphere is not conserved during the burst. The
correct conserved quantity, which we have considered in this section,
is the angular momentum per particle $L$. In the slow rotation
approximation, our result (eq.~[\ref{eq:slow}]) agrees with equation
(9) of Abramowicz et al.~(2001).

\begin{figure}
\epsscale{1.0}\plotone{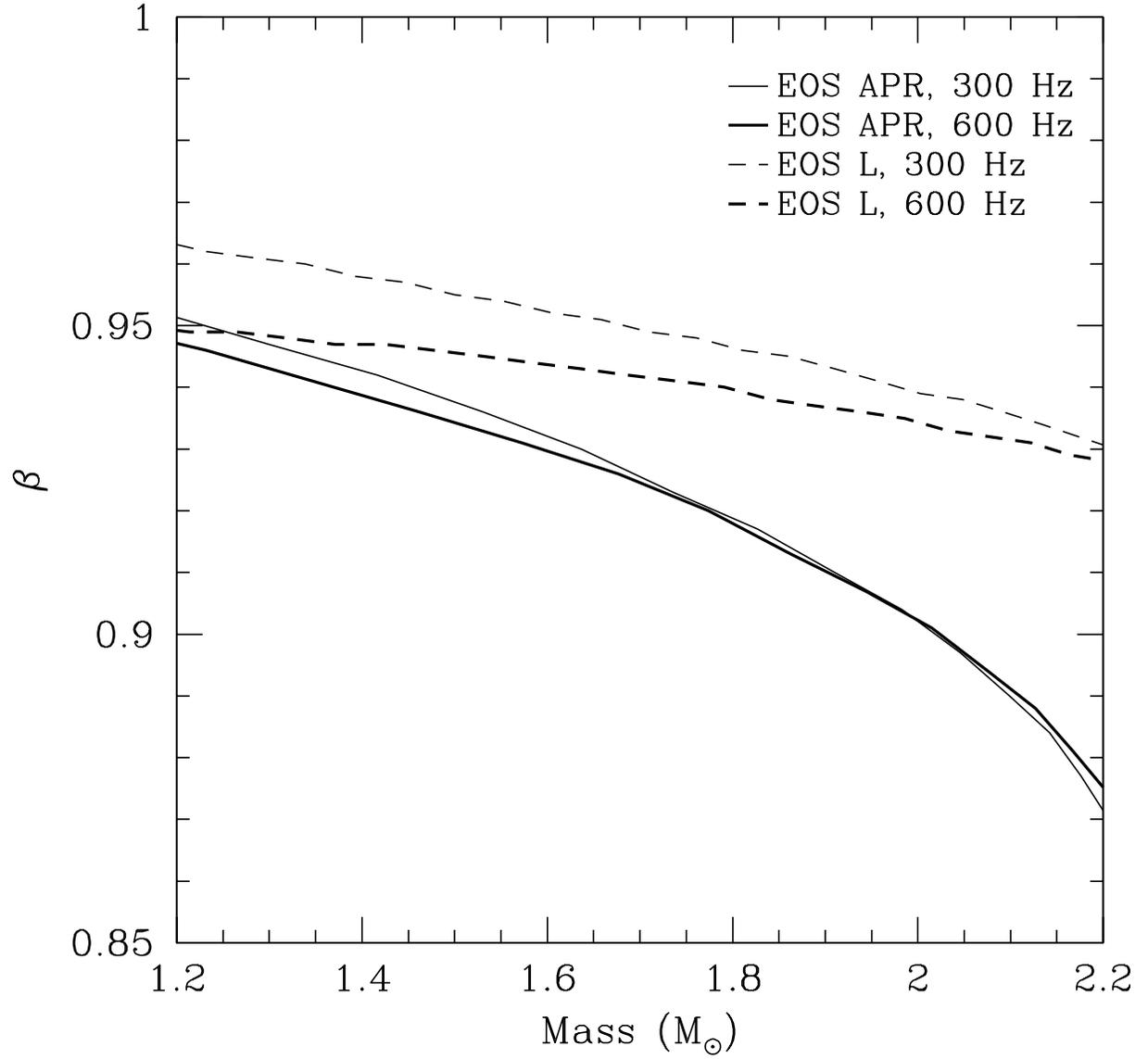}\epsscale{1.0}
\caption{The general relativistic correction to the Newtonian angular
momentum conservation law. We plot $\beta=(-1/2)(d\ln\Omega/d\ln r)$
as a function of neutron star mass for EOS L ({\it dashed curves}) and
EOS APR ({\it solid curves}), and for $\nu=300\ {\rm Hz}$ ({\it light
curves}) and $\nu=600\ {\rm Hz}$ ({\it heavy
curves}).\label{fig:beta}}
\end{figure}

We write $d\ln\Omega/d\ln r$ as $-2\beta$, where
\begin{equation}
\beta\equiv -{1\over 2}{d\ln\Omega\over d\ln r}
\end{equation}
is unity in the Newtonian limit. We give the value of $\beta$ in Table
\ref{tab:models}. Figure \ref{fig:beta} shows $\beta$ as a function of
neutron star mass for EOS APR and EOS L, and for $\nu=300$ and $600\
{\rm Hz}$. We see that including the general relativistic angular
momentum conservation law changes $d\ln\Omega/d\ln r$ by 5--10\%.

\section{Calculation of Expansion and Spin Down}

In this section, we present calculations of the expansion and
spin-down of the atmosphere, including the effects of general
relativity and rapid rotation. In the spirit of CB, we ignore
latitudinal variations, and work in the equatorial plane. First, in \S
4.1, we discuss the effects of general relativity on the equations
describing the hydrostatic and thermal structure of the atmosphere,
and calculate the reduction in gravity at the equator due to rapid
rotation. In \S 4.2, we show how to include the general relativistic
angular momentum conservation law that we obtained in \S 3. In \S 4.3,
we rescale our fiducial results of \S 2 for the spin-down of a
rigidly-rotating atmosphere to different masses, equations of state,
and spin frequencies. In \S 4.4, we relax the assumption of
rigid-rotation, and present the rotational profiles of the atmosphere
for several different cases.

\subsection{Hydrostatic and Thermal Structure of the Atmosphere}

We now write down the differential equations describing the
hydrostatic and thermal structure of the atmosphere including general
relativity, and compare them to the Newtonian equations. We start by
considering a non-rotating star, and then discuss the additional
effects of rotation.

For a spherical star, Thorne (1977) (see also Thorne 1967) gives the
equation for mass conservation as
\begin{equation}\label{eq:mass}
{dM_r\over dr}=4\pi r^2\rho{\mathcal V}(r),
\end{equation}
where $M_r$ is the rest mass (number of baryons multiplied by baryon
rest mass) within coordinate radius $r$, $\rho$ is the rest mass
density, and ${\mathcal V}$ is the ``volume redshift factor''. For
slow rotation, ${\mathcal V}(r)=(1-2M(r)/r)^{-1/2}$, where $M(r)$ is
the gravitational mass interior to $r$. We adopt a plane parallel
approximation in the thin envelope, and write $r=R+z/{\mathcal V}$,
where $z\ll R$ is the proper length. Here, ${\mathcal V}$ is the
redshift factor at the surface of the star ${\mathcal V}={\mathcal
V}(R)$. We define the rest mass column depth $dy=-dM_r/4\pi R^2$. The
mass conservation equation becomes
\begin{equation}\label{eq:1}
{dz\over dy}=-{1\over\rho}.
\end{equation}
The equation for hydrostatic balance is (Thorne 1977)
\begin{equation}\label{eq:HB}
{dP\over dM_r}=-{GM(r)\over 4\pi r^4}{\mathcal V},
\end{equation}
where $P$ is the locally measured pressure, and we neglect
contributions to the gravitational mass from the fluid energy
density. Using the definition of column depth, we rewrite equation
(\ref{eq:HB}) for the thin layer as
\begin{equation}\label{eq:2}
{dP\over dy}=g,
\end{equation}
where
\begin{equation}\label{eq:g}
g={GM\over R^2}{\mathcal V}.
\end{equation}
Rewriting the entropy and flux equations (Thorne 1977, eqs. [11d] and
[11h]) in a similar way, we find
\begin{equation}\label{eq:3}
{dF\over dy}=-\epsilon
\end{equation}
and
\begin{equation}\label{eq:4}
{dT\over dy}={3\kappa\over 4acT^3}F,
\end{equation}
where $F$ is the heat flux, $\epsilon$ is the nuclear energy release
rate per unit rest mass, $\kappa$ is the opacity, and $T$ is the
locally measured temperature. Equations (\ref{eq:1}), (\ref{eq:2}),
(\ref{eq:3}), and (\ref{eq:4}), which describe the structure of the
atmosphere, are identical to the Newtonian equations (compare CB \S
2).

These results also hold in the slow rotation approximation, for which
the star is spherical to first order in $(\Omega/\Omega_K)^2$. The
inclusion of rapid rotation has three effects. The first is to change
the volume correction factor ${\cal{V}}$. In addition, the mass
conservation equation must be covariant so that all observers agree on
the rest mass of the star. This requirement gives an additional
Lorentz factor in equation (\ref{eq:mass}) (e.g., Friedman et
al.~1986). However, we find that for the models we consider in this
paper, both of these effects are $\lesssim 1$\%, and so we neglect
them.

Much more important is the correction to gravity due to the rapid
rotation. This correction factor is the relativistic generalisation of
the Newtonian reduction in gravitational acceleration due to the
centrifugal force. To calculate this, we write the four-velocity as
$u^\alpha = {\cal{R}}(t^\alpha + \Omega \phi^\alpha)$, and use the
normalization $u^\alpha u_\alpha = -1$ to solve for the redshift
factor
\begin{equation}
{\cal{R}} = e^{-(\gamma+\rho)/2}\frac{1}{\sqrt{1-v^2}},
\end{equation}
where $v$ is given by equation (\ref{eq:v}). The equation of
hydrostatic balance for a rigidly rotating star is then (Friedman et
al 1986)
\begin{equation}
\frac{1}{\rho} \frac{dP}{dr} = - \frac{d}{dr} {\log \cal{R}},
\end{equation}
where we assume the pressure is much smaller than the rest mass
density. To reduce this equation to the form of equation (\ref{eq:2}),
we write
\begin{equation}\label{eq:g2}
g = \frac{GM}{R^2} {\cal V A},
\end{equation}
where the correction factor $\cal{A}$ accounts for the reduction in
$g$ at the equator due to rapid rotation,
\begin{equation}
{\cal A} = \frac{R^2}{GM {\cal V}^2} \frac{d}{dr} {\log \cal{R}},
\end{equation}
tending to unity in the limit of zero rotation. This is the most
important effect coming from rotation, giving a correction of up to
25\% for a rotation frequency of $600\ {\rm Hz}$ and a stiff equation
of state.

In summary, the general relativistic equations look the same as the
Newtonian equations if we use the correct value of surface gravity $g$
(eq.~[\ref{eq:g2}]), and if we identify the column depth $y$ with the
rest mass column depth, and the thickness $z$ with the proper
thickness of the layer. Values of $g$, ${\mathcal V}$, and ${\cal A}$
are given in Table \ref{tab:models} for different models.

\subsection{Calculation of Spin Down}

To incorporate the general relativistic angular momentum conservation
law into our calculations of spin down, we write the angular momentum
per unit rest mass at the equator in the form
\begin{equation}\label{eq:f}
u_\phi=\Omega f(r),
\end{equation}
where $f(r)$ is a function of radius, e.g.~$f(r)=r^2$ in the Newtonian
limit. If the angular momentum is constant, we find that the change of
$\Omega$ with radius is determined by the gradient of $f$ with radius:
$\beta=(-1/2)(d\ln\Omega/d\ln r)=(1/2)(d\ln f/d\ln r)$. Now consider a
particle which moves from radius $R$ to $R+\Delta r$, where $\Delta
r\ll R$. Conservation of angular momentum gives $u_\phi(R+\Delta
r)=u_\phi(R)$. Expanding in a Taylor series around $r=R$, we find
\begin{equation}
{\Delta\Omega\over\Omega}=-\beta\left({2\Delta r\over R}\right).
\end{equation}
In the Newtonian limit, for which $\beta=1$, we recover the familiar
result $\Delta\Omega/\Omega=-2\Delta r/R$. To calculate the moment of
inertia, we must modify equation (\ref{eq:right}). Again writing the
angular momentum in the form of equation (\ref{eq:f}), and expanding
$f(r)$ around $r=R$, we find
\begin{equation}
I={8\pi R^2f(R)\over 3}\int \left[1+{2\beta z(y)\over {\mathcal V}R}\right]\ dy,
\end{equation}
which reduces to equation (\ref{eq:right}) in the Newtonian limit.

\subsection{Scaling our Fiducial Results to Different Masses
and Equations of State}

The calculations presented in \S 2 were for a surface gravity
$g=1.9\times 10^{14}\ {\rm cm\ s^{-2}}$ (the Newtonian gravity for a
$1.4 M_\odot$, $R=10\ {\rm km}$ neutron star). We now rescale these
fiducial results to different masses and equations of state, including
the effects of rapid rotation and general relativity.

To calculate the thickness of the atmosphere in \S 2, we integrated
hydrostatic balance in the form $dP/dz=-\rho\,GM/R^2$, giving the
thickness $\Delta z/R\propto 1/gR\propto R/M$ (see also \S 2 of CB). In
addition, we must include a factor ${\cal V}$ to obtain the correct
general relativistic value for $g$, and a factor ${\cal A}$ to include
the effects of rotation on gravity. We thus find the fractional change
in thickness of the atmosphere is
\begin{equation}
{\Delta z\over R}={\mathcal V}^{-1}{\mathcal A}^{-1}\left({0.21\over
M/R}\right)\left({\Delta z\over R}\right)_0,
\end{equation}
where $(\Delta z/R)_0$ is the fiducial value from \S 2. The fractional
change in spin frequency is $\beta(2\Delta r/R)$, where $\Delta r$ is
the coordinate thickness. Converting to proper thickness $\Delta z$
involves another factor of ${\cal V}$, giving
\begin{equation}
{\Delta\Omega\over\Omega}=\beta{\mathcal V}^{-2}{\mathcal
A}^{-1}\left({0.21\over M/R}\right)
\left({\Delta\Omega\over\Omega}\right)_0\equiv
\alpha\left({\Delta\Omega\over\Omega}\right)_0,
\end{equation}
where we define $\alpha$ as the scaling factor from the fiducial
value, and we took $M/R=0.21$ in \S 2. Note that the largest
contributions to $\alpha$ come from rescaling gravity and including
the centrifugal force, rather than from general relativistic effects.

\begin{figure}
\epsscale{1.0}\plotone{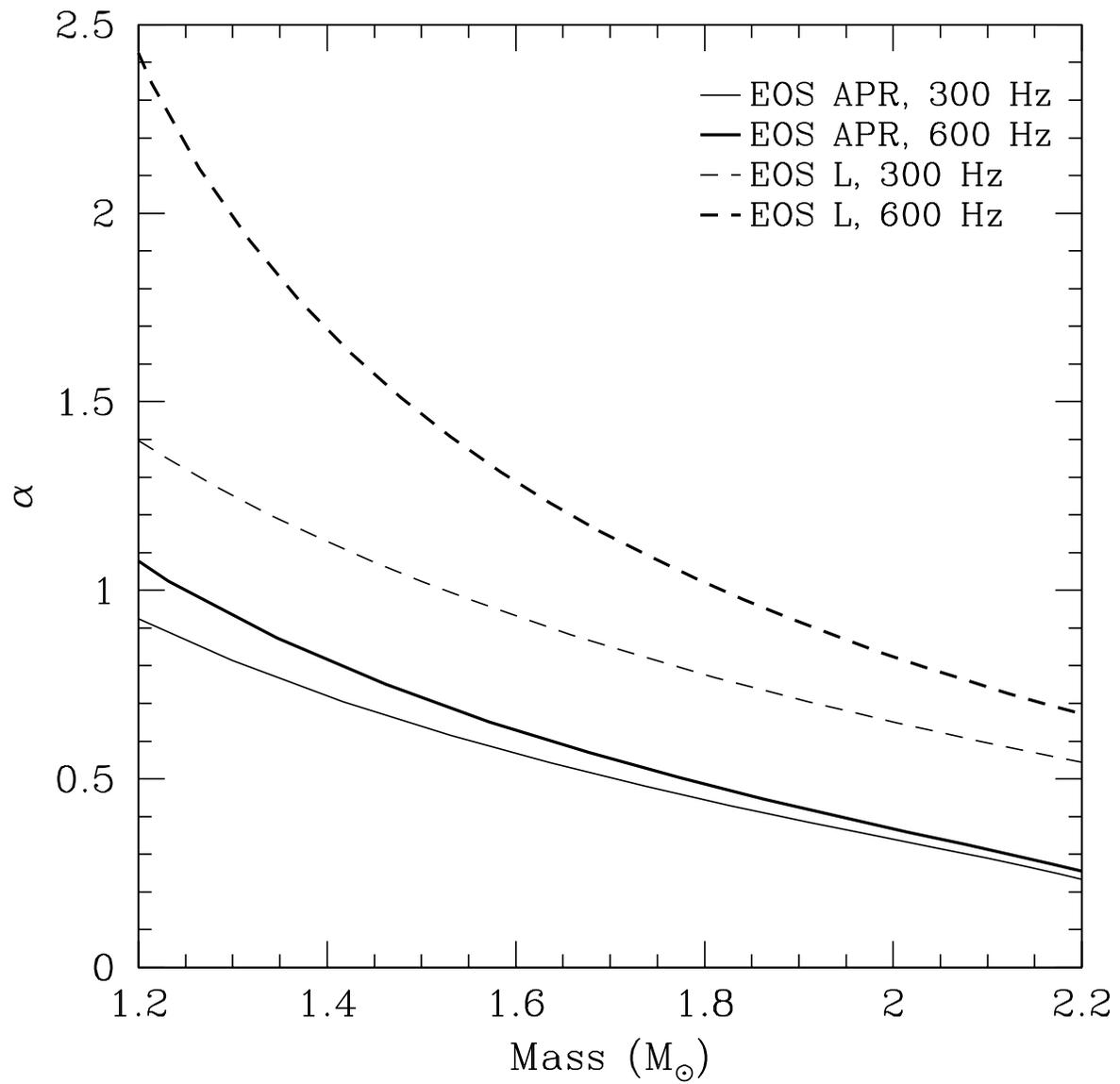}\epsscale{1.0}
\caption{The scaling factor $\alpha$ which rescales the fiducial
values of $\Delta\Omega/\Omega$ presented in \S 2 and Figure 1 to
different neutron star masses, equations of state, and spin
frequencies. The factor $\alpha$ includes the scaling with surface
gravity, the correction to gravity at the equator due to rapid
rotation, and general relativistic corrections to gravity and the
angular momentum conservation law. We show results for EOS L ({\it
dashed curves}) and EOS APR ({\it solid curves}), and for $\nu=300\
{\rm Hz}$ ({\it light curves}) and $\nu=600\ {\rm Hz}$ ({\it heavy
curves}).\label{fig:corr}}
\end{figure}

Figure \ref{fig:corr} shows $\alpha$ as a function of neutron star
mass, for EOS APR and EOS L, and in each case for $\nu=300$ and $600\
{\rm Hz}$. We also give $\alpha$ in Table \ref{tab:models}. Figure
\ref{fig:corr} shows that for massive neutron stars with $M\approx 2
M_\odot$, the frequency shifts calculated in \S 2 and shown in Figure
1 should be multiplied by $0.65$--$0.85$ for EOS L, and $0.3$--$0.4$
for EOS APR. For a $1.4 M_\odot$ star, these factors are $1.1$--$1.7$
for EOS L, and $0.7$--$0.8$ for EOS APR. The largest values of
$\alpha$ are obtained for a low neutron star mass, stiff equation of
state, and rapid rotation. However, even $\alpha=1.7$ for $M=1.4\
M_\odot$, EOS L, and $\nu=600\ {\rm Hz}$ is not large enough to give
agreement with the largest observed frequency shifts.

\subsection{Rotational Profiles}

\begin{figure}
\epsscale{1.0}\plottwo{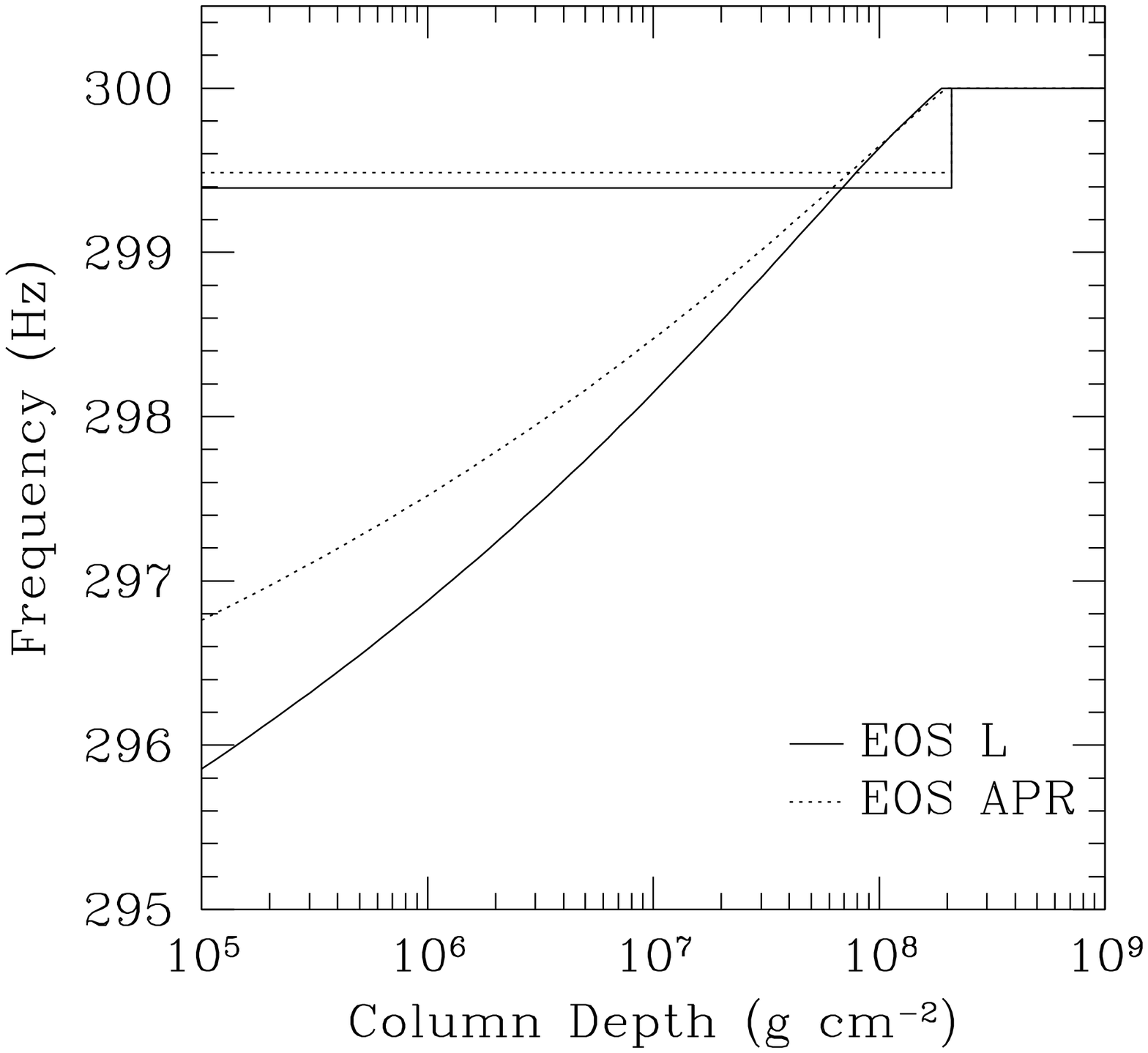}{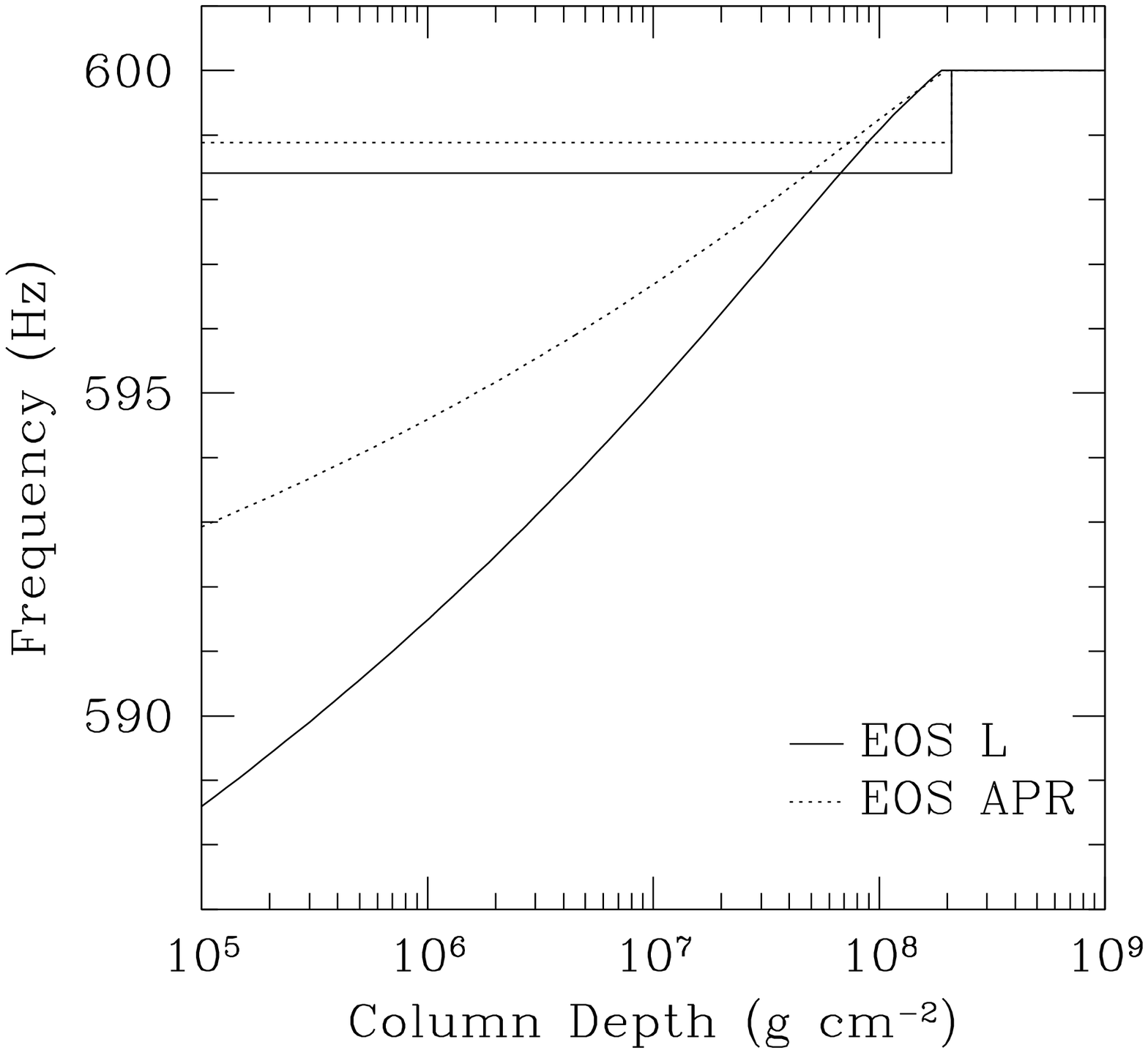}\epsscale{1.0}
\caption{Detailed rotational profiles for $1.4\ M_\odot$ neutron stars
undergoing mixed H/He ignition at a global accretion rate of $0.1\
\dot M_{\rm Edd}$. We assume that the atmosphere is rigidly-rotating
immediately prior to the burst, and that during the burst, the
atmosphere is radiative and carries a flux equal to the
solar-composition Eddington flux at the photosphere. For $\nu=300\
{\rm Hz}$ ({\it left panel}) and $\nu=600\ {\rm Hz}$ ({\it right
panel}), and for EOS L ({\it solid lines}) and EOS APR ({\it dotted
lines}), we plot the rotational profiles assuming either (i) complete
rotational coupling, resulting in rigid rotation across the layer, or
(ii) no angular momentum transport, giving a differentially-rotating
layer.
\label{fig:omega}}
\end{figure}

In Figure \ref{fig:omega}, we present some rotational profiles for
some specific models. We take the neutron star mass to be $1.4\
M_\odot$, the global accretion rate to be $0.1\ \dot M_{\rm Edd}$, and
again work in the equatorial plane. We assume that the atmosphere is
rigidly-rotating immediately prior to the burst, and that during the
burst, the atmosphere is radiative and carries a flux equal to the
solar-composition Eddington flux at the photosphere ($F_{\rm
Edd}=8.8\times 10^{24}\ g_{14}\ {\rm erg\ cm^{-2}\ s^{-1}}$). For
$\nu=300$ and $\nu=600\ {\rm Hz}$, and for EOS L and EOS APR, we plot
the rotational profiles assuming either (i) complete rotational
coupling, resulting in rigid rotation across the layer, or (ii) no
angular momentum transport, giving a differentially-rotating layer.

If we allow differential rotation to persist, we see that the upper
layers of the atmosphere spin down by an amount comparable to or
greater than the observed spin changes during bursts. However, it is
not clear why the spin frequency observed should be that of only the
outermost layers. Indeed, CB (\S 3.4) argued that substantial
differential rotation would wash out any signal from the deeper
cooling layers because of the finite time to transport heat
vertically.

%==================================================================
\section{Summary and Discussion}

We have presented new calculations of the hydrostatic expansion and
spin-down of a neutron star atmosphere during a Type I X-ray
burst. Our main conclusion is that hydrostatic expansion is not enough
to explain the observed frequency drifts during Type I X-ray bursts if
the burning atmosphere rotates rigidly.

In \S 2, we showed that Cumming \& Bildsten (2000) (CB) overestimated
the change in the moment of inertia of the atmosphere, obtaining
values of spin-down that were a factor of two too large. Figure 1
compares the new calculations of spin-down with observations,
including recent measurements of large frequency drifts during bursts
(Galloway et al.~2000; Wijnands et al.~2001). We find that the largest
observed frequency shift is a factor of 3 or more greater than the
theoretical values.

In \S 3, we derived the angular momentum conservation law in general
relativity. We calculated the variation of spin frequency with radial
distance, $d\ln\Omega/d\ln r$, for a particle moving with constant
angular momentum. In the slow rotation approximation ($\Omega^2\ll
GM/R^3$), we obtained the analytic result given by equation
(\ref{eq:slow}), which agrees with recent work by Abramowicz et
al.~(2001). For rapidly-rotating stars, we calculated $d\ln\Omega/d\ln
r$ by numerically solving for the structure of the neutron star. The
correction to the Newtonian angular momentum conservation law,
$\beta=(-1/2)(d\ln\Omega/d\ln r)$, is shown in Figure 2 for different
neutron star masses, equations of state and spin frequencies. Contrary
to the results of Heyl (2000), which were also shown to be incorrect
by Abramowicz et al.~(2001), we find that the general relativistic
correction is small, about 5--10\%.

In \S 4, we calculated the atmospheric expansion and spin-down,
including the effects of rapid rotation and general
relativity. Working in the equatorial plane (in the spirit of CB's
calculation, we neglect latitudinal variations in this paper), we
calculated the scaling factor $\alpha$ required to rescale our
fiducial results of \S 2 to different neutron star masses, equations
of state, and spin frequencies (Figure 3). In addition, we presented
the rotational profiles for some particular models (Figure 4). We find
that the largest spin-down is for rapid rotation, and low mass stars
with a stiff equation of state. For example, a 1.4 $M_\odot$ star
spinning at 600 Hz with the stiff equation of state EOS L has
$\alpha=1.7$. However, this is not a large enough factor to bring
observations and theory into agreement in Figure 1.

The rotational profiles given in Figure 4 show that in principle
frequency shifts as large as those observed can be obtained by
hydrostatic expansion, if we consider only the outermost layers of the
atmosphere, and allow differential rotation. However, it is not at all
obvious why the observed frequency would be that of the outermost
shells of the atmosphere, which contain a small amount of the mass,
particularly since the energy release is in the deeper layers. Indeed,
CB argued (see their \S 3.4) that substantial differential rotation in
the burning layers would wash out the signal, because of the finite
time needed to transport heat vertically in the atmosphere.

Another possible objection to the angular momentum conservation
picture was given by CB, who pointed out that if the burning layers
are threaded by a large scale poloidal magnetic field, this field will
be wound up by the differential rotation during the burst. The wound
up toroidal field acts back on the shear, halting and reversing its
direction in the time for an Alfven wave to cross the atmosphere (see,
for example, Spruit 1999). For a magnetic field typical of a
millisecond radio pulsar, $B\sim 10^8\ {\rm G}$, this timescale is
only $\sim 0.01\ {\rm s}$. One possibility is that the surface field
is much weaker, $B\lesssim 10^6\ {\rm G}$, allowing shearing to
persist. Observations of burst oscillations in the accreting
millisecond X-ray pulsar SAX~J1808.4-3658, which perhaps has a $\sim
10^8$--$10^9\ {\rm G}$ field (Psaltis \& Chakrabarty 1999), would give
an interesting test of this picture. In't Zand et al. (2000) report a
marginal detection with BeppoSAX of oscillations at $400\pm 2\ {\rm
Hz}$ during a Type I burst, but are unable to resolve any frequency
drift. Unfortunately, as yet no burst oscillations have been detected
from this source with RXTE.

The results we have obtained in this paper suggest that we may have to
look elsewhere for an explanation of the frequency drifts. Spitkovsky,
Levin, \& Ushomirsky (2001), in a detailed investigation of
hydrodynamic flows during Type I bursts, propose that the frequency
drifts may be explained by a combination of radial expansion and
variations in the velocity of zonal flows on the neutron star
surface. Another possibility mentioned by CB is that the burst
oscillation is due to a non-radial oscillation in the neutron star
surface layers. Such possibilities remain to be investigated in
detail.

\acknowledgements

We thank Jeremy Heyl and Scott Hughes for many useful discussions, and
Yuri Levin for a careful reading of the manuscript. We are grateful to
Chris Matzner for insights into the issues discussed in the
Appendix. This work began at the program on Spin and Magnetism in
Young Neutron Stars, ITP, Santa Barbara. A. C. and S. M. thank the
Canadian Institute for Theoretical Astrophysics for hospitality during
its completion. This research was supported by the National Science
Foundation under grants PHY99-07949 and AY97-31632, by NASA via grant
NAG 5-8658, and by the Natural Sciences and Engineering Research
Council of Canada. L. B. is a Cottrell Scholar of the Research
Corporation.

\begin{deluxetable}{llllcllll}
\tablecaption{Neutron Star Models\label{tab:models}}
\tablewidth{0pt}
\tablehead{
\colhead{$M\ (M_\odot)$} & \colhead{$R\ ({\rm km})$}
& \colhead{$M/R$} & \colhead{${\mathcal V}$\tablenotemark{a}}
& \colhead{$g\ (10^{14}\ {\rm cm\ s^{-2}})$} & \colhead{${\mathcal A}$\tablenotemark{b}}   & \colhead{$\beta$\tablenotemark{c}}  &
\colhead{$\alpha$\tablenotemark{d}} &
\colhead{$\nu_B$\tablenotemark{e}$\ ({\rm Hz})$}}
\startdata
\cutinhead{ EOS APR $\nu = 300$ Hz}
1.4 & 11.5 & 0.18 & 1.25 & 1.72 & 0.98 & 0.94 & 0.72 & 1070\\  
2.0 & 11.1 & 0.27 & 1.46 & 3.10 & 0.99 & 0.90 & 0.34 & 1330\\  
\cutinhead{ EOS APR $\nu = 600$ Hz}
1.4 & 11.9 & 0.17 & 1.24 & 1.48 & 0.91 & 0.94 & 0.81 & 1070\\  
2.0 & 11.3 & 0.26 & 1.45 & 2.83 & 0.95 & 0.90 & 0.37 & 1330\\  
\cutinhead{ EOS L $\nu = 300$ Hz}
1.4 & 15.2 & 0.14 & 1.17 & 0.90 & 0.95 & 0.96 & 1.13 & 730\\  
2.0 & 15.3 & 0.19 & 1.28 & 1.39 & 0.97 & 0.94 & 0.65 & 850\\  
\cutinhead{ EOS L $\nu = 600$ Hz}
1.4 & 16.7 & 0.12 & 1.15 & 0.55 & 0.72 & 0.95 & 1.70 & 730\\  
2.0 & 16.2 & 0.18 & 1.25 & 1.05 & 0.83 & 0.93 & 0.82 & 850\\  
\enddata
\tablenotetext{a}{The redshift factor ${\mathcal V}=(1-2M/R)^{-1/2}$ (we neglect the
$\lesssim 1$\% correction due to rotation).}
\tablenotetext{b}{The correction to gravity due to
rapid rotation, $g=GM{\mathcal VA}/R^2$.}
\tablenotetext{c}{The general relativistic correction to the angular
momentum conservation law, $\beta\equiv (-1/2)(d\ln\Omega/d\ln r)$.}
\tablenotetext{d}{The multiplicative scaling factor that should be
applied to the results of \S 2 and Fig.~1, to allow for the different
gravity, and the effects of general relativity and rapid rotation (see
text).}
\tablenotetext{e}{The break-up spin frequency.}
\end{deluxetable}

\clearpage
\appendix
\section{Angular Momentum Conservation for a Fluid}

In this Appendix, we start with the conservation equations for a
fluid, which include contributions from the fluid's enthalpy and from
heat flow, and show that the appropriate conserved quantity is the
angular momentum per baryon $u_\phi$ as calculated in \S 3.

We begin with the stress tensor for a fluid
\begin{equation}
T^{\alpha\beta}=h\rho u^\alpha u^\beta+Pg^{\alpha\beta}+F^\alpha u^\beta+
u^\alpha F^\beta,
\end{equation}
(see, for example, Mihalas \& Mihalas 1984). The enthalpy $h$ is
\begin{equation}\label{eq:h}
h={\epsilon+P\over\rho}
\end{equation}
where the energy density
\begin{equation}\label{eq:eps}
\epsilon=\rho c^2+\rho \Pi - \rho B
\end{equation}
is the sum of rest mass energy, internal energy, and nuclear binding
energy. We write the heat flux as $F^\alpha$. By setting
$\phi_\beta\nabla_\alpha T^{\alpha\beta}=0$, and using baryon number
conservation $\nabla_\alpha (\rho u^\alpha) = 0$, we obtain the
angular momentum conservation law
\begin{equation}\label{eq:a1}
\rho u^\alpha\nabla_\alpha\left(hu_\phi+{F_\phi\over\rho}\right)+
\nabla_\alpha\left(F^\alpha u_\phi\right)=0.
\end{equation}
Assuming the azimuthal heat flux $F^\phi$ may be neglected when
compared to the vertical heat flux $F^r=F$, we rewrite equation
(\ref{eq:a1}) as
\begin{equation}\label{eq:a2}
{d\over dt}\left(hu_\phi\right)=-{1\over\rho}{d\over
dz}\left(Fu_\phi\right).
\end{equation}
We now simplify this equation using the first law of thermodynamics,
which is
\begin{equation}\label{eq:firstlaw}
d\epsilon=\left({\epsilon+P\over\rho}\right)d\rho+\rho Tds
\end{equation}
where $s$ is the entropy per baryon mass, or, using equation
(\ref{eq:h}),
\begin{equation}
dh=Tds+{dP\over\rho}.
\end{equation}
Substituting for $dh/dt$ in equation (\ref{eq:a2}), we find
\begin{equation}\label{eq:a3}
{du_\phi\over dt}=-{u_\phi\over h}\left[T{ds\over dt}+{1\over
\rho}{dF\over dz}\right] -{u_\phi\over\rho h}{dP\over dt}-{F\over\rho
h}{du_\phi\over dz}.
\end{equation}
The first term in equation (\ref{eq:a3}) vanishes, since
$Tds/dt=-(1/\rho)dF/dz$ is the familiar entropy equation. The nuclear
energy production rate is included in the $Tds/dt$ term, since the
nuclear binding energy is included in the enthalpy:
equations (\ref{eq:eps}) and (\ref{eq:firstlaw}) give
\begin{equation}
T{ds\over dt}={d\Pi\over dt}-{P\over\rho^2}{d\rho\over dt}-{dB\over
dt},
\end{equation}
where $dB/dt$ is the nuclear energy generation rate. Integrating the
second term in equation (\ref{eq:a3}) in time, we find it gives a
contribution to $\Delta u_\phi/u_\phi$ of order $\Delta P/\rho
c^2\approx (H/R)(P/\rho c^2)\approx (gR/c^2)(H/R)^2\approx 10^{-7}$,
where we take the pressure scale height $H\approx 10\ {\rm m}$. The
third term in equation (\ref{eq:a3}) represents radiative viscosity
(see Mihalas \& Mihalas 1984). Again integrating in time, we find that
this term gives a contribution $\approx (F/\rho c^2)(t/H)(\Delta
u_\phi/u_\phi)\approx 10^{-4}(t/{\rm sec})(\Delta
u_\phi/u_\phi)$. Both the second and third terms are of order
$(H/R)^2$ or smaller, and may be neglected for our purposes.

In summary, consideration of the conservation law for a fluid gives
\begin{equation}
{du_\phi\over dt}=0+{\mathcal O}\left({H\over R}\right)^2,
\end{equation}
showing that, at the level of accuracy we require, the angular
momentum per baryon $u_\phi$ is conserved.

\end{document}